\RequirePackage{fix-cm}
\documentclass{article}
\usepackage{graphicx}
\usepackage[affil-it]{authblk}
\usepackage{amsmath}
\usepackage{amsfonts}
\usepackage{epstopdf}

\date{}
\begin{document}

\title{Memoization technique for optimizing functions with stochastic input \thanks{Supported by Ministry of Education of Republic Serbia (project III 044006)}}

\author[1]{Edin H. Mulali\'c}
\author[2]{Miomir S. Stankovi\'c}
\author[3]{Radomir S. Stankovi\'c}
\affil[1]{Mathematical Institute of Serbian Academy of Sciences and Arts, Kneza Mihaila 36, 11001 Belgrade, Serbia}
\affil[2]{University of Ni\v{s}, Faculty of Occupational Safety, \v{C}arnojevi\'{c}a 10A, 18000 Ni\v{s}, Serbia}
\affil[3]{University of Ni\v{s}, Faculty of Electronic Engineering, Aleksandra Medvedeva 14, 18000 Ni\v s, Serbia}

\maketitle

\begin{abstract}
In this paper we present a strategy for optimization functions with stochastic input. The main idea is to take advantage of decomposition in combination with a look-up table. Deciding what input values should be used for memoization is determined based on the underlying probability distribution of input variables. Special attention is given to difficulties caused by combinatorial explosion.

\textbf{Keywords:} optimization, dynamic programming, functional decomposition, stochastic input, look-up table, memoization

\end{abstract}

\section{Introduction}
\label{intro}
A basic operation such as calculating a value of a function is in the heart of most problem solving processes. In specialized systems (particularly in military and other real time systems) where the speed of calculation is of great importance and one particular function is a bottleneck, various optimization techniques could be applied. There is no general recipe for successful optimization. It usually requires problem dependant heuristic, for example:
	\begin{itemize}
		\item different ways of representing the function
		\item hardware implementation or combination of hardware and software implementation
		\item different algorithm
		\item improved data structures used in the algorithm
		\item decomposition
		\item parallelism
		\item pre computed values
		\item approximative solutions
	\end{itemize}
	In this paper, we will explore usage of additional resource in optimizing functions with stochastic input. We will see how to take advantage of functional decomposition and present a solution for resource allocation. Several algorithms based on dynamic programming are used to deal with the problem of combinatorial explosion. We will also make an attempt to keep the story as general as possible. The rest of the paper is organized as follows. Section \ref{sec:definition} introduces terminology used through the paper. In this section, the problem is presented in a formal way and several key questions to be answered are emphasised. Section \ref{sec:solutionMain} discusses the proposed solution of previously presented problem. Section \ref{sec:solutionFurther} gives additional details of the proposed solution and optimization technique for the critical step in the solution. And, finally, section \ref{sec:conclusion} gives conclusion of the presented work.

\section{Problem definition}
\label{sec:definition}
Let's suppose that we have given a finite commutative ring $ (\mathbf{U}, \oplus, \otimes) $, where $ \mathbf{U} = \{u_1, u_2, ..., u_K\}, $ $ K \in  \mathbb{N}. $ We want to evaluate function $ f:\mathbf{U}^N \to \mathbb{R}, $ $ N \in \mathbb{N} $. We can write $ f(\mathbf{x}) = f(x_1, x_2, ..., x_N) $ where $ x_i \in \mathbf{U} $ for $ 1 \leq i \leq N. $ Computing function value for any specific input vector requires time $ T_c(f). $ Let's assume that we have additional resource of limited size M units. We can use one or more resource units to optimize calculation of function value for one or more input vectors.

If we have a memory of limited size $ M, $ we would be able to pre compute and store function values for up to M values of input parameter combinations (input vectors). Assuming that reading a value from the memory requires constant time $ T_M $ and that $ T_M $ is significantly less than $ T_c(f), $ with this approach we can cut down the average time of evaluating the function $ f. $ Note that the term \textit{memory} in this context can denote a physical memory or a convenient data structure. Look-up tables were popular in the world of mathematics even before invention of modern computers. Such tables were used mostly to avoid manually calculating complex functions (trigonometry functions or logarithms, for example) \cite{campbell2003history}. In computer science, using look-up tables have become standard optimization technique in many areas. In designing logical circuits, look-up tabls are used because of speed and flexibility, since changing software is much easier than changing hardware. In computer programming, \textit{memoization} is a well-known technique to avoid repeating calculations. In particular, within a system, a function could be invoked multiple times with the same input arguments. Therefore, it would be useful to store computed values, and compute from scratch only for those values of input parameters not seen before. Although this technique is often used by programmers, manually implementing such mechanism often requires significant changes in source code and could be tedious and time-consuming. That is the reason for some programming environments to provide automated memoization \cite{hall1997improving}. 

If we don't know anything about input variables $ x_i, $ random $ M $ combinations of inputs (out of $ K^N $) could be used for pre-computing. But what if input parameters are not of deterministic nature? If the input parameters have stochastic nature, obviously we can use better strategy for selecting $ M $ the most useful input combinations. Let's assume that there is an underlying probability distribution, so that probability of $ x_i = u_j $ is denoted as $ p_{ij}, $ where $ \sum_{j=1}^{j=K} p_{ij} = 1, $ for each $ i, 1 \leq i \leq N. $ We will also assume that input parameters have independent distributions. Those distributions could be known in advance, before designing the system. Alternatively, distributions of input parameters could be learned on-line, during the work of a system which implements the function. Information about those distributions could be used to find $ M $ \textit{most probable }combinations of input parameters and use them as precomputed and stored values. Obviously, that will minimize the expected time of evaluating the function which is given by formula:
 \begin{align*}
	E_f[T] &= \sum_{\mathbf{x}^{\prime} \in X_M}P(\mathbf{x}^{\prime})T_M + \sum_{\mathbf{x}^{\prime} \notin X_M}P(\mathbf{x}^{\prime})T_c(f)	\\
		 &= T_MP(X_M) + T_c(f)(1 - P(X_M))	\\ 
		 &= T_c(f) - P(X_M)(T_c(f) - T_M),
\end{align*} 
where
\begin{itemize}
	\item $ X_M $ is the set of all input vectors used for pre-computation,
	\item $ P(\mathbf{x}^{\prime}) $ is the probability that an input vector is $ \mathbf{x}^{\prime}, $
	\item $ P(X_M) $ is the probability that an input vector belongs to the set $ X_M. $
\end{itemize}
Of course, this is not the only way of using the memory resource. If we store function values for $ M $ input vectors, we basically did two things. First, we reduced average evaluation time. Second, we significantly improved calculation for those $ M $ vectors. But for all other $ K^N-M $ input vectors, evaluation time is still $ T_c $. Depending on the usage of the system, this might be satisfying solution. But there are some issues in this approach. First, is it possible to use memory resource in a different way to reduce average evaluation time even more? And second, how can we affect more than $ M $ input vectors? One way to approach these two problems is functional decomposition.

\section{Functional decomposition and optimal resource distribution}
\label{sec:solutionMain}
%We start this section by giving a general definition of functional decomposition. 
%\begin{definition}
\textbf{Definition 1.} A decomposition $ \Delta(f) $ of a function $ f $ is set of functions $ \Delta(f) = \{F, f_1, f_2 ..., f_D\}, $ such that
\begin{align*}
	f(\mathbf{x}) &= f(x_1, x_2, ...x_N) \\
				  &= F(f_1(\mathbf{x_1}), f_2(\mathbf{x_2}), ..., f_D(\mathbf{x_D}))
\end{align*}
where components of each vector $ \mathbf{x_i} (1 \leq i \leq D) $ are from the set of components of the initial vector $ \mathbf{x} $. 
%\end{definition}
Decomposition traditionally plays important role in many areas of mathematics and computer science. It is in the heart of problem solving strategy ``divide and conquer'' and has been particularly significant in the areas where parallelism is of great importance. It remains one of the key problems in logic synthesis \cite{voudouris2005decomposition} ever since Ashenhurst \cite{ashenhurst1957decomposition}, Curtis \cite{curtis1962new}, Roth and Carp \cite{roth1962minimization} did pioneering work in this field, but it has also important applications in many other fields of engineering \cite{selvaraj2006functional}. When combined with look-up tables, decomposition is a powerful tool for representation of a function in a more economical way.
%\begin{example}

\textbf{Example 1} Let's suppose that we have given function 
\[
h(x_1, x_2, x_3, x_4, x_5, x_6) = x_1x_2x_3 + x_4x_5x_6,
\]
where $x_i \in \left\lbrace 0, 1 \right\rbrace.$ Representing the function $h$ in memory would require storing $2^6 = 64$ values. Now, let's suppose that we decomposed the function $h$ in the following way:
\[
h(x_1, x_2, x_3, x_4, x_5, x_6) = h_1(x_1, x_2, x_3) + h_2(x_4, x_5, x_6),
\]
where
\[
h_1(x_1, x_2, x_3) = x_1x_2x_3, \quad  h_2(x_4, x_5, x_6) = x_4x_5x_6.
\]
Representing functions $h_1$ and $h_2$ would require storing $2^3 + 2^3 = 16$ values in total. So the function $h$ could be calculated from those $16$ values with the price of one additional operation $+$.   
%\end{example}
To optimize a function $f(\textbf{x})$ by using a look-up table (of total size $M$) in combination with a decomposition $\Delta(f) = F(f_1(\mathbf{x_1}), f_2(\mathbf{x_2}), ..., f_D(\mathbf{x_D}))$, we must find the optimal distribution of the memory resource among available functions $f_i(\mathbf{x_i})$. The first step in finding an optimal resource distribution is to calculate average time for evaluating the function $ f. $ The expected time is given by the following formula:
\begin{eqnarray}
\label{equ: Expected time}
	E_{f, M, \Delta}[T] &=& T_c(\Delta(f)) - \sum_{j=1}^D P(X_M^{(j)} | m_j)(T_c(f_j(\mathbf{x_j})) - T_M)
\end{eqnarray}
where
\begin{itemize}
  \item $T_c(\Delta(f))$ is time of calculating the function $f$ in the decomposition form $\Delta(f)$ without using the additional resources,
  \item $P(X_M^{(j)} | m_j)$ is the probability that output of the function $f_j$ can be obtained from memory without calculation under the condition that function $f_j$ has available $m_j$ memory locations,
  \item $T_c(f_j(\mathbf{x_j}))$ is the time of calculating the function $f_j$ without using the additional resources.
\end{itemize}
In order to minimise expected time from equation (\ref{equ: Expected time}), we are looking to maximise
\begin{equation}
\label{equ:to maximise}
	\sum_{j=1}^D P(X_M^{(j)} | m_j)(T_c(f_j(\mathbf{x_j})) - T_M) = \sum_{j=1}^D \omega_j(m_j),
\end{equation}
where
\[
\omega_j(m_j) = P(X_M^{(j)} | m_j)(T_c(f_j(\mathbf{x_j})) - T_M).
\]
So, resource allocation problem can be defined as: find values for $m_j, (1 \leq j \leq D)$ for which the expression $\sum_{j=1}^D \omega_j(m_j)$ is maximised while the following conditions are satisfied:
\begin{equation}
\label{cond: m inequality}
	0 \leq m_j \leq M
\end{equation}
\begin{equation}
\label{cond: m sum}
	\sum_{j=1}^D m_j \leq M
\end{equation}
One way to solve this problem is by brute force - for each combination $[m_1...m_D]$ which satisfies conditions given by equations (\ref{cond: m inequality}) and (\ref{cond: m sum}), calculate the expression given by the equation (\ref{equ:to maximise}) and find the best among them. The problem with this algorithm is its exponential complexity. By following procedure outlined in \cite{dreyfus1977art}, we obtain the algorithm based on dynamic programming which solves the problem in polynomial time.
Let's define matrix $ A[i, j] $ as
\begin{equation*}
A[i, j] = \operatorname{max} \sum_{k=1}^j \omega_k(m_k),
\end{equation*}
where
\begin{equation*}
\sum_{k=1}^j m_k = i.
\end{equation*}
We will also introduce variable $l_j$ as the length (number of components) of vector $\mathbf{x_j}$. The following procedure gives the desired solution.

\textit{Initialization.} For $i=0,...,l_1$
\begin{align*}
A(i, 1) &= \omega_1(i)	\\
B(i, 1) &= 0
\end{align*}
\textit{Recursion.} For $j=2,...,D$ and $i=0,...,\operatorname{min} \{\sum_{k=1}^j K^{l_k}, M \}$
\begin{equation*}
A(i, j+1) = \underset{i^{\prime}, \operatorname{max}\{0, i-K^{l_{j+1}}\} \leq i^{\prime} \leq i}{\operatorname{max}} [A(i^{\prime}, j) + \omega_{j+1}(i-i^{\prime})]
\end{equation*}
\begin{equation*}
B(i, j+1) = \underset{i^{\prime}, \operatorname{max}\{0, i-K^{l_{j+1}}\} \leq i^{\prime} \leq i}{\operatorname{argmax}} [A(i^{\prime}, j) + \omega_{j+1}(i-i^{\prime})]
\end{equation*}
\textit{Stopping and reconstruction.}
\begin{equation*}
i^*_D = \underset{i, 0 \leq i \leq \operatorname{min}\{\sum_{k=D}^j K^{l_k}, M \}}{\operatorname{argmax}} A[i, D]
\end{equation*}
For $ j = D-1, D-2, ..., 1 $
\begin{align*}
i_j^* &= B[i^*_{j+1}, j+1]	\\
m_1 &= i^*_1
\end{align*}
For $ j = 2, 3.., D $
\begin{equation*}
m_j = i^*_j - i^*_{j-1}
\end{equation*}
Time complexity of the described procedure is $ O(M^2D) $ and space complexity is $ O(MD). $

\section{Efficiently calculating $\omega_j(m_j)$}
\label{sec:solutionFurther}

%\begin{figure*}
%  \includegraphics[width=0.75\textwidth]{v.png}
%\caption{Finding k-best...}
%\end{figure*}

In the previously described algorithm, calculating $\omega_j(m_j)$ can be tricky. Brute force algorithm gives exponential complexity, so once again the dynamic programming can be helpful. Here, the main challenge lies in the calculation of values $ P_{ij} = P(X_M^{(j)} | m_j=i) $ for each $i$, $ 0 \leq i \leq L$, $L = \operatorname{min} \{M, K^{l_j}\}$ and finding the corresponding $i$ vectors for which the output of the function should be memorized. All those values can be calculated simultaneously by reducing the problem to finding $L$-best paths in trellis. By following procedure from \cite{seshadri1994list}, we design an algorithm which solves the problem in polynomial time. First, we will present the algorithm for finding $ P(X_M^{(j)} | m_j=1) $ (i.e. single best path in a trellis) and then we will adapt the algorithm to find values $ P(X_M^{(j)} | m_j=i) $ for each valid $i$ (i.e. $L$-best paths).

\subsubsection{Finding the best path in a trellis}
Let's $\mathbf{x_j} = (y_1, y_2, ..., y_{N_y}),$ $y_i \in \mathbf{U}$ for $i=1,...,N_{N_y}$. Let $ \Psi_t(i) $ be the probability of the most probable vector of length $t$ (or equivalently the most probable path of length $t$) $(y_1, y_2, ..., y_t)$ for which $y_t=u_i$. For such vector, let's $\epsilon_t(i)$ be the value of the component $y_{t-1}$. The following algorithm gives most probable vector and its probability.

\textit{Initialization ($ t=1 $)} For $ 1 \leq i \leq K$
\begin{align*}
	\Psi_t(i) &= p(y_1 = u_i)	\\
	\epsilon_t(i) &= 1
\end{align*}
\textit{Recursion ($ 1 < t \leq N_y $)} For $ 1 \leq i \leq K $
\begin{equation*}
	\Psi_t(i) = \underset{1 \leq j^{\prime} \leq K}{\operatorname{max}} [\Psi_{t-1}(j^{\prime}) * p(y_t = u_i)]
\end{equation*}
\begin{equation*}
	\epsilon_t(i) = \underset{1 \leq j^{\prime} \leq K}{\operatorname{argmax}} [\Psi_{t-1}(j^{\prime}) * p(y_t = u_i)]
\end{equation*}
\textit{Stopping and reconstruction.} The probability of the most probable vector is given by
\begin{equation*}
	P^* = \underset{1 \leq j^{\prime} \leq K}{\operatorname{max}} [\Psi_{N_y}(j^{\prime})].
\end{equation*}
The most probable vector is given by
\begin{equation*}
	(y^*_{N_y}) = \underset{1 \leq j^{\prime} \leq K}{\operatorname{argmax}} [\Psi_{N_y}(j^{\prime})],
\end{equation*}
and for $ t = N_y-1, N_y-2...1, $
\begin{equation*}
	y^*_t = \epsilon(y^*_{t+1}).
\end{equation*}

\subsubsection{Finding $L$-best paths in a trellis}

Let's $ \Psi_t(i, k) $ be the probability of the $k$-th most probable vector (or equivalently the $k$-th most probable path in trellis) $(y_1, y_2, ..., y_t)$ for which $y_t=u_i$. For such vector, let's $\epsilon_t(i, k)$ be the value of component $y_{t-1}.$ The following algorithm gives $L$ most probable vectors and their probabilities. Complexity of the described procedure is $O(kK^2N_y)$.

\textit{Initialization ($ t=1 $)} For $ 1 \leq i \leq K, 1 \leq k \leq L $
\begin{align*}
	\Psi_t(i, k) &= p(y_1 = u_i)	\\
	\epsilon_t(i, k) &= 1
\end{align*}
\textit{Recursion ($ 1 < t \leq N_y $)} For $ 1 \leq i \leq K $
\begin{align*}
	\Psi_t(i, k) = \underset{1 \leq j^{\prime} \leq K, 1 \leq l^{\prime} \leq L}{\operatorname{max^{(k)}}} [\Psi_{t-1}(j^{\prime}, l^{\prime}) * p(y_t = u_i)],	\\
(j^*, l^*) = \underset{1 \leq j^{\prime} \leq K, 1 \leq l^{\prime} \leq L}{\operatorname{argmax^{(k)}}} [\Psi_{t-1}(j^{\prime}, l^{\prime}) * p(y_t = u_i)],
\end{align*}
\begin{align*}
	\epsilon_t(i, k) &= j^*	\\
	r_t(i, k) &= l^*
\end{align*}
where $\operatorname{max^{(k)}}$ denotes $k$-th largest value.

\textit{Stopping and reconstruction.} Probability of the $k$-th most probable vector is given by
\begin{equation*}
	P^*_k = \underset{1 \leq j^{\prime} \leq K, 1 \leq l^{\prime} \leq L}{\operatorname{max^{(k)}}} [\Psi_{N_y}(j^{\prime}, l^{\prime})]
\end{equation*}
Now, it is easy to obtain $k$-th most probable vector:
\begin{equation*}
	(y^*_{N_y}, l^*_{N_y}) = \underset{1 \leq j^{\prime} \leq K, 1 \leq l^{\prime} \leq L}{\operatorname{argmax^{(k)}}} [\Psi_{N_y}(j^{\prime}, l^{\prime})]
\end{equation*}
and for $ t = N_y-1, N_y-2,...,1, $
\begin{align*}
	y^*_t &= \epsilon(y^*_{t+1}, l^*_{t+1})	\\
	l^*_t &= r(y^*_{t+1}, l^*_{t+1})
\end{align*}

The value $P(X_M^{(j)}|m_j=i)$ can be calculated by summing the $i$ best probabilities obtained by the previously described algorithm.

\section{Conclusion}
\label{sec:conclusion}

Stochastic signals appear often in real life. Functions which input has stochastic nature are common in real-time systems. In this paper, we have described one possible technique for optimization that type of functions. It is based on using additional memory resources for speeding up the calculation of functions for certain values of input vectors. We proposed dynamic programming procedure which can determine the optimal resource distribution for a particular decomposition in polynomial time. By using proposed procedure, one could compare different decompositions and pick the best one, but discovering various decompositions was outside of the scope of this paper. However, it has been one of the hot topics in the science and is certainly an interesting problem for future work.

\end{document}